\documentclass{PoS}

\makeatletter
\def\gsim{\compoundrel>\over\sim}
\def\lsim{\compoundrel<\over\sim}
\def\compoundrel#1\over#2{\mathpalette\compoundreL{{#1}\over{#2}}}
\def\compoundreL#1#2{\compoundREL#1#2}
\def\compoundREL#1#2\over#3{\mathrel
	{\vcenter{\hbox{$\m@th\buildrel{#1#2}\over{#1#3}$}}}}
\makeatother
\newcommand{\bfi}[1]{\mbox{\boldmath $#1$}}
\usepackage{graphicx}

\title{
Central and tensor Lambda-Nucleon potentials from lattice QCD
}

\ShortTitle{
Central and tensor $\Lambda N$ potentials from lattice QCD
}

\author{\speaker{Hidekatsu Nemura}%
	 \\
	 Department of Physics,
	 Tohoku University,
	 Sendai,
	 980-8578, Japan \\
	E-mail: \email{nemura@nucl.phys.tohoku.ac.jp}}

\author{
        for HAL QCD and PACS-CS Collaboration\\
	\includegraphics[width=0.20\textwidth]{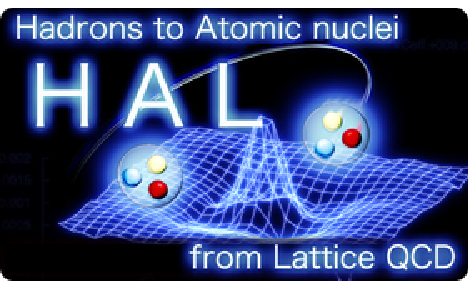}
	}

\abstract{
We present our latest study of Lambda-Nucleon ($\Lambda N$) interaction
by using lattice QCD, following up on our report at 
LATTICE 2008. %
We have calculated not only the scattering lengths but also the central and 
tensor potentials, which are obtained from the
Bethe-Salpeter (BS) amplitude measured in lattice QCD.
For these calculations, we employ two different types of gauge configurations:
	  (i) 2+1 flavor full  QCD configurations generated by the PACS-CS collaboration
	  at $\beta=1.9$ ($a=0.0907(13)$ fm) on a $32^3\times 64$ lattice, whose spatial volume is  
	  (2.90 fm)$^3$, with the quark masses corresponding to 	
	  $(m_\pi,m_K)\approx (301,592)$, %
	  $(414,637)$, %
	  $(570,724)$ %
	  and
	  $(699,787)$ (in units of MeV). %
	  (ii) Quenched QCD configurations at $\beta=5.7$ ( $a=0.1416(9)$ fm)
	  on a $32^3\times48$ lattice, 
	   whose spatial volume is  
	  (4.5 fm)$^3$, with the quark masses corresponding to 	 
	  $(m_\pi,m_K)\approx (512,606)$, %
	  $(464, 586)$ %
	  and
	  $(407,565)$. %
The following qualitative features are found:
The $\Lambda p$ potential has a relatively strong (weak) repulsive
core in the $^1S_0$ ($^3S_1$) channel at short distance,
while the potential has slight attractive region at medium distance. 
The tensor potential is found to be weaker than the $NN$ case.
These results hold in both full and quenched QCD. %
The energy of the ground state on the finite lattice volume is calculated:  
In both spin channels, energy shift due to the finite volume, from which we extract the scattering length
via the L\"{u}scher's formula, is found to be
negative  at all values of  quark masses,  
suggesting that the $\Lambda N$ interaction is attractive.
We have also discussed the quark mass dependences of the potentials and the scattering lengths.
}

\FullConference{The XXVII International Symposium on Lattice Field Theory - LAT2009\\
		 July 26-31 2009\\
		 Peking University, Beijing, China}

\begin{document}

\section{Introduction}

The hyperon-nucleon ($YN$) and the hyperon-hyperon ($YY$) 
interactions are the keys to explore the strange nuclear
systems, in which hyperons (or strange quarks) are embedded 
in normal nuclear systems as ``impurities''~\cite{Hashimoto:2006aw}. 
For example,  
spectroscopic studies of the $\Lambda$ and $\Sigma$ hypernuclei are 
carried out experimentaly and theoretically, which leads to 
a qualitative conclusion that the $\Lambda$-nucleus interaction 
is attractive while the $\Sigma$-nucleus interaction is repulsive.
Such information is useful to study 
 the composition of hyperonic matter inside the 
 neutron stars~\cite{SchaffnerBielich:2008kb}:
the $\Lambda$ particle instead of $\Sigma^-$ 
 would be the first strange baryon to appear in   the core of the 
  neutron stars.
The $\Xi N$ interaction is also interesting and important,  
in order to explore the existence of $\Xi$ hypernuclei 
and the dense hyperonic  matter in neutron stars. 
Despite their importance, 
 $YN$ and $YY$ interactions have still large uncertainties
  because direct $YN$ and $YY$
   scattering experiments are either difficult or 
   impossible due to the short life-time of hyperons. 

Under these circumstances,  
the lattice QCD would be a valuable theoretical tool to
make a first-principle calculation of baryon-baryon interactions.
 Previously,  scattering parameters 
based on the L\"{u}scher's formula have been
reported for the $NN$ system \cite{Fukugita:1994ve,Beane:2006mx} and 
for the $YN$ system \cite{Muroya:2004fz, BeaneYN2007}.
Recently, a new approach to the $NN$ 
interaction from the lattice QCD has been proposed \cite{Ishii:2006ec}. 
In this approach,  the $NN$ potential can be directly obtained from lattice QCD
through the Bethe-Salpeter (BS) amplitude and the observables
 such as the phase shift and the binding energy  can be calculated
  by using the resultant potential. 
 See Refs.~\cite{
Aoki:2008wy,MuranoProcLat09,%
Nemura0806.1094,%
InoueProcLat09,%
IshiiProcLat09,%
Aoki:2009ji%
} 
for the recent developments in various aspects 
along the line of this approach. 

The purpose of this report is to present 
our recent calculation of the 
$\Lambda N$ potentials 
together with the scattering parameters 
by using the full and quenched QCD gauge configurations. 
A preliminary result was already reported 
at LATTICE 2008\cite{Nemura:2009kc}.
This is the latest version of the report which includes  
several new efforts:
(i) Not only the central potential but also the tensor potential
are calculated,
(ii) quark mass dependences of the potentials and 
the scattering lengths are studied, and 
(iii) statistics of the lattice are increased from the previous report.

\section{Formulation}

The basic formulation has already been given in 
Refs.~\cite{Ishii:2006ec,Nemura0806.1094,Aoki:2008hh,Ishii:2009zr}.
(See also Refs.~\cite{Luscher,CPPACS.PRD71_94504_2005}.)
and a recent comprehensive accounts for the lattice $NN$ potential is 
found in \cite{Aoki:2009ji}. 
 The formulation for the $YN$ interaction is reported in
  \cite{Nemura:2009kc}, so that we mention only the basic equations
  briefly below.  
We start from an effective Schr\"{o}dinger equation for the
equal-time BS wave function:
\begin{equation}
 -{1\over 2\mu}\nabla^2 \phi(\vec{r}) +
  \int U(\vec{r},\vec{r}^\prime)
  \phi(\vec{r}^\prime) d^3r^\prime  =
  E \phi(\vec{r}).
\end{equation}
Here $\mu=m_{\Lambda}m_{N}/(m_{\Lambda}+m_{N})$ and 
$E\equiv k^2/(2\mu)$ are the reduced mass of the $\Lambda N$ system and 
the non-relativistic energy in the center-of-mass frame, respectively. 
We consider the low-energy scattering state so that the 
nonlocal potential can be rewritten 
 by the derivative or velocity expansion~\cite{TW67},
$
U(\vec{r},\vec{r}^\prime)=
 V_{\Lambda N}(\vec{r},\vec{\nabla})\delta(\vec{r}-\vec{r}^\prime).
$
The general expression of the potential $V_{\Lambda N}$ 
is known to be~\cite{JJdeSwart1971} 
\begin{eqnarray}
 V_{\Lambda N} &=&
  V_0(r)
  +V_\sigma(r)(\vec{\sigma}_{\Lambda}\cdot\vec{\sigma}_{N})
  +V_T(r)S_{12}
  +V_{LS}(r)(\vec{L}\cdot\vec{S}_+)
  +V_{ALS}(r)(\vec{L}\cdot\vec{S}_-)
  +{O}(\nabla^2).
  \label{GenePotNL}
\end{eqnarray}
Here
$S_{12}=3(\vec{\sigma}_{\Lambda}\cdot\vec{n})(\vec{\sigma}_{N}\cdot\vec{n})-\vec{\sigma}_{\Lambda}\cdot\vec{\sigma}_{N}$
is the tensor operator with $\vec{n}=\vec{r}/|\vec{r}|$,
$\vec{S}_{\pm}=(\vec{\sigma}_{N} \pm \vec{\sigma}_{\Lambda})/2$  are %
symmetric ($+$) and antisymmetric ($-$) spin operators,
$\vec{L}=-i\vec{r}\times\vec{\nabla}$ is the orbital %
angular momentum operator. %
 $V_{0,\sigma,T}$ are the leading order (LO) potentials
 while $V_{LS,ALS}$ are the next-to-leading-order (NLO) potentials
 in the velocity expansion.
The procedure to derive the LO potentials is as follows.
For the spin triplet state, %
the noncentral forces (e.g. the tensor force) 
mix different partial waves such as $S$- and $D$-waves. 
Namely, 
the spin-triplet wave function $\phi(r;J=1)$ comprises 
the $S$- %
and 
the $D$-wave components, %
which can be
extracted from the lattice wave function $\phi_{\alpha\beta}(r;J=1)$
such that
(See \cite{Nemura:2009kc} for the method to obtain $\phi_{\alpha\beta}(r,J)$
in lattice QCD.) 
\begin{equation}
 \left\{
 \begin{array}{l}
  \phi_{\alpha\beta}(r;\ ^3S_1)={\cal P}\phi_{\alpha\beta}(r;J=1)
   \equiv {1\over 24} \sum_{{\cal R}\in{ O}} {\cal R}
   \phi_{\alpha\beta}(r;J=1),
   \\
  \phi_{\alpha\beta}(r;\ ^3D_1)={\cal Q}\phi_{\alpha\beta}(r;J=1)
   \equiv (1-{\cal P})\phi_{\alpha\beta}(r;J=1).
 \end{array}
 \right.%
\end{equation}
Therefore, 
the effective Schr\"{o}dinger equation with the LO 
potentials becomes: 
\begin{equation}
 \left\{
 \begin{array}{c}
  {\cal P} \\
  {\cal Q}
 \end{array}
 \right\}
 \times
 \left\{
  -{1\over 2\mu}\nabla^2 
  +V_0(r)
  +V_\sigma(r)(\vec{\sigma}_{\Lambda}\cdot\vec{\sigma}_{N})
  +V_T(r)S_{12}
 \right\}
 \phi(\vec{r})
 =
 \left\{
 \begin{array}{c}
  {\cal P} \\
  {\cal Q}
 \end{array}
 \right\}
 \times
 E \phi(\vec{r})
\end{equation}
The $r$ dependence of the central and the tensor potentials, 
$V_C(r;J=0)=V_0(r)-3V_\sigma(r)$ for $J=0$, 
$V_C(r;J=1)=V_0(r) +V_\sigma(r)$, and $V_T(r)$ for $J=1$, are 
determined once 
we obtain 
the wave function, the total  energy and the reduced mass 
in lattice QCD.

\section{Numerical simulations}

\subsection{2+1 flavor  QCD}

Main results in this report are obtained by using 
the 2+1 flavor full QCD gauge configurations generated by PACS-CS collaboration~\cite{Aoki:2008sm}
with  the RG-improved Iwasaki gauge action and the nonperturbatively $O(a)$-improved 
Wilson quark action at $\beta=1.9$ on a $32^3\times 64$ lattice.
The lattice spacing at the physical quark masses has been estimated as
$a=0.0907(13)$ fm~\cite{Aoki:2008sm}.  
So far, 
we have employed four values of the hopping parameter for light quarks, 
$\kappa_{ud}= 0.13700, 0.13727, 0.13754, 0.13770$, 
while the one for the strange quark is fixed to $\kappa_s= 0.13640$. 
Several light hadron masses obtained in the present calculation 
are shown in Table~\ref{masses}.
To calculate the BS wave function, 
the wall source is placed at the time-slice $t_0$ with  the Coulomb gauge fixing, 
and the Dirichlet boundary condition is imposed 
in the temporal direction at the time-slice $t-t_0=32$. 
In order to improve the statistics, 
multiple sources at $t_0=8n$ with $n=0,1,2,\cdots, 8$
are employed on each gauge configuration. 
The results are obtained 
with $N_{\rm conf}=609, 481, 568, 422$   
for $\kappa_{ud}= 0.13700, 0.13727, 0.13754, 0.13770$, respectively,
where $N_{\rm conf}$ is the number of the gauge configurations.

\subsection{Quenched QCD with larger spatial volume
  }

In quenched QCD calculation, 
we employ the plaquette gauge action and the
Wilson quark action at $\beta=5.7$ on a $32^3\times 48$ lattice.
The periodic boundary condition is imposed for
quarks in the spatial direction. 
The wall source is placed at $t_0=0$ 
with the Coulomb gauge fixing and the 
Dirichlet boundary condition is imposed at $t=24$ in the temporal direction. 
The lattice spacing 
at the physical point 
is determined as 
$a=0.1416(9)$~fm ($1/a = 1.393(9)$ GeV)
from $m_\rho=770$ MeV.  The spatial lattice volume is $(4.5\mbox{fm})^3$. 
The hopping parameter for the strange quark mass 
is given by $\kappa_s=0.16432(6)$ from 
$m_K=494$ MeV. 
Three values of the hopping parameter, 
$\kappa_{ud}=0.1665, 0.1670, 0.1675$,
are employed for the light quark mass. 
Table~\ref{masses} also lists the light hadron masses calculated 
in quenched QCD. 
The results in quenched QCD are obtained with  
$N_{\rm conf}\approx1000$.

\begin{table}[b]
 \centering \leavevmode 
 \begin{tabular}{ccccccccc}
  \hline \hline
  $\kappa_{ud}$
  & 
  $m_\pi$ & $m_\rho$ & $m_K$ & $m_{K^\ast}$ & 
  $m_N$ & $m_\Lambda$ & $m_{\Sigma}$ & $m_{\Xi}$ \\
  \hline
  \multicolumn{9}{l}{\bf 2+1 flavor QCD by PACS-CS with
  ${\bf \kappa_{\it s}=0.13640}$
  } \\
  $0.13700$
  &
  699.4(4) & 
  1108(3) &
  786.8(4) &
  1159(2) &
  1572(6) &
  1632(4) &
  1650(5) &
  1701(4) \\
  $0.13727$
  &
  567.9(6) & 
  1000(4) &
  723.7(7) &
  1081(3) &
  1396(6) &
  1491(4) &
  1519(5) &
  1599(4) \\
  $0.13754$
  &
  413.6(6) & 
  902(3) &
  636.6(4) &
  1026(3) &
  1221(7) &
  1349(4) &
  1406(8) &
  1505(4) \\
  $0.13770$
  &
  301(3) & 
  845(10) &
  592(1) &
  980(6) &
  1079(12) &
  1248(15) &
  1308(13) &
  1432(7) \\
  \hline
  \multicolumn{9}{l}{\bf quenched QCD with
  ${\bf \beta=5.7, \kappa_{\it s}=0.1643}$
  } \\
  $0.1665$ &
  511.8(5)  & 862(3)  & 605.8(5)  & 898(1) &
  1297(6) & 1344(6) & 1375(5) & 1416(3) \\
  $0.1670$ &
  463.6(6) & 842(4)  & 586.3(5) & 895(2) &
  1250(9)  & 1314(9) & 1351(6)  & 1404(4) \\
  $0.1675$ &
  407(1)   & 820(3)  & 564.9(5) & 886(3) &
  1205(13) & 1269(9) & 1326(9)  & 1383(5) \\
  \hline
  \hline
  Exp. & 
  135   &  770   & 494  & 892   &
  940   & 1116   & 1190 & 1320  \\
 \hline \hline 
  \end{tabular}
 \caption{
 Hadron masses in the unit of MeV. 
 }
 \label{masses}
\end{table}

\section{Numerical results}

\subsection{2+1 flavor QCD}

\begin{figure}[t]
 \centering \leavevmode
  \includegraphics[width=.90\textwidth]
  {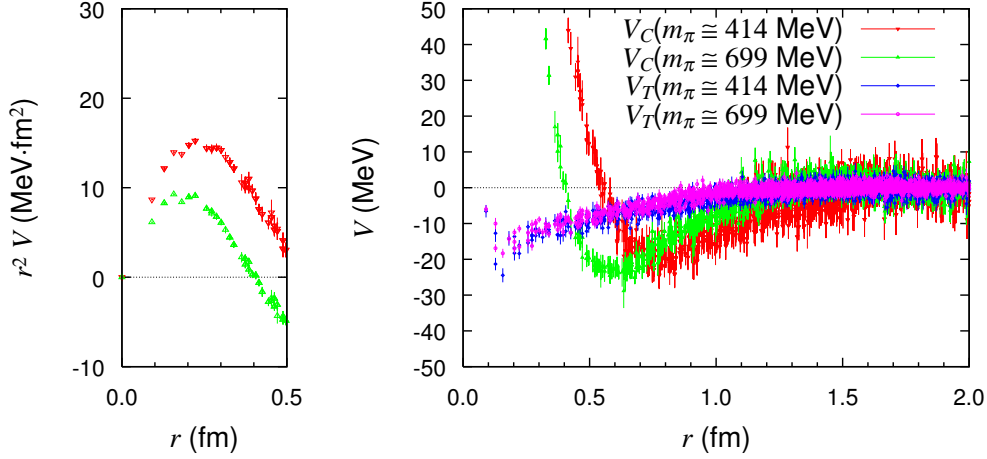}%
 \caption{
  The central and the tensor potentials in $^3S_1-^3D_1$ channel
 in $2+1$ flavor QCD as a function of $r$ at $m_\pi\simeq 414$ MeV (red and blue) and 699 MeV (green and magenta).
 }
 \label{PACSCS_VCT_3E1}
\end{figure}

\begin{figure}[h]
 \centering \leavevmode
  \includegraphics[width=.90\textwidth]
  {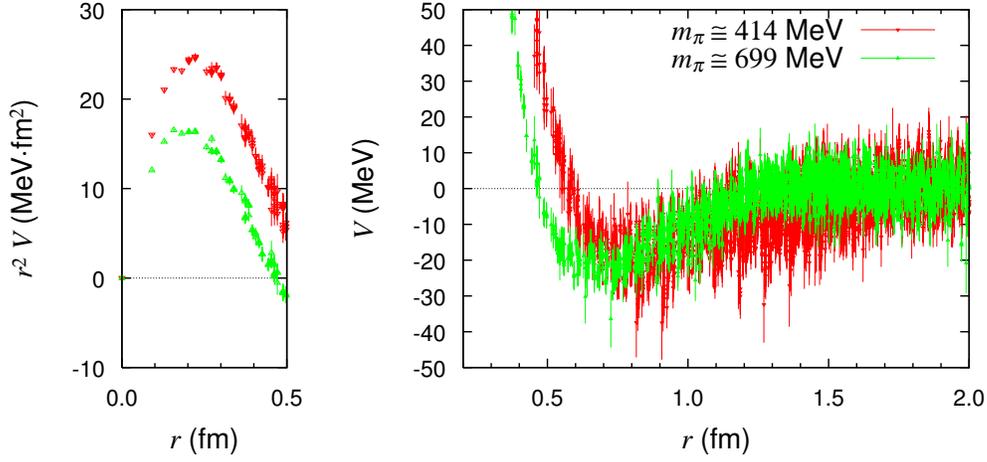}%
 \caption{
 The central potential in $^1S_0$ channel in $2+1$ flavor QCD as a function of $r$ at $m_\pi\simeq 414$ MeV (red) and 699 MeV (green).
 }
 \label{PACSCS_VCeff_1S0}
\end{figure}

Figures~\ref{PACSCS_VCT_3E1} and~\ref{PACSCS_VCeff_1S0} 
show  the $\Lambda N$ potentials 
obtained from 2+1 flavor QCD calculation as a function of $r$.
The central ($V_C(J=1)$) and the tensor ($V_T$) potentials in 
the $^3S_1-^3D_1$ channels are given in Fig.~\ref{PACSCS_VCT_3E1} 
while the central potential in the $^1S_0$ channel ($V_C(J=0)$) 
is given in Fig.~\ref{PACSCS_VCeff_1S0}. 
We also show the central potential multiplied by volume 
factor ($r^2 V_C(r)$) in the left panel 
in addition to the normal $V(r)$ given in the right panel, 
in order to compare the strength of the repulsive force between 
two quark masses. 
These figures contain results with
$(m_\pi,m_K) \approx (699,787)$ and $(414,637)$ MeV,
which are obtained at  $t-t_0=13$ and $10$, respectively. 
These time-slices are chosen so that the ground state saturation 
is achieved.

As can be seen in both figures, 
the attractive well of the central potential 
moves to outer region as the $u,d$ quark mass decreases 
while the depth of these attractive pockets do not change 
so much. 
The present results show that the tensor force is weaker than the 
$NN$ case
 \cite{Ishii:2009zr}, and 
the quark mass dependence of the tensor force seems to be small. 
Both of the repulsive and attractive parts increase in magnitude 
as the $u,d$ quark mass decreases.

For $m_\pi \approx 700$ MeV, the central potentials 
reach $V_C \rightarrow 0$ at the radial distance
$r \sim 1.3$ fm,
which is smaller than the half of the physical lattice
length ($aL/2 \approx 1.45$ fm). 
Therefore the L\"{u}scher's formula can be applied to extract 
the scattering phase shift, which will be discussed in the 
latter subsection. 
For $m_\pi \approx 400$ MeV, on the other hand, 
the interaction range of the $V_C$, which is about $1.4$ fm, 
almost reaches to the half of the lattice. 
Therefore we must be very careful to extract the scattering 
phase shift at this or lighter quark masses from the 
L\"{u}scher's formula, 
though no sign of the violation against the L\"{u}scher's 
condition was observed within errors for the effective 
central potential even at $m_\pi \approx 300$ MeV. 
(See Fig.~1 in the previous report~\cite{Nemura:2009kc}.)
Calculation on larger spatial volume will be needed to 
correctly extract the scattering phase shift 
at $m_\pi \approx 300$ MeV.


\begin{figure}[t]
 \centering \leavevmode
  \includegraphics[width=.90\textwidth]
  {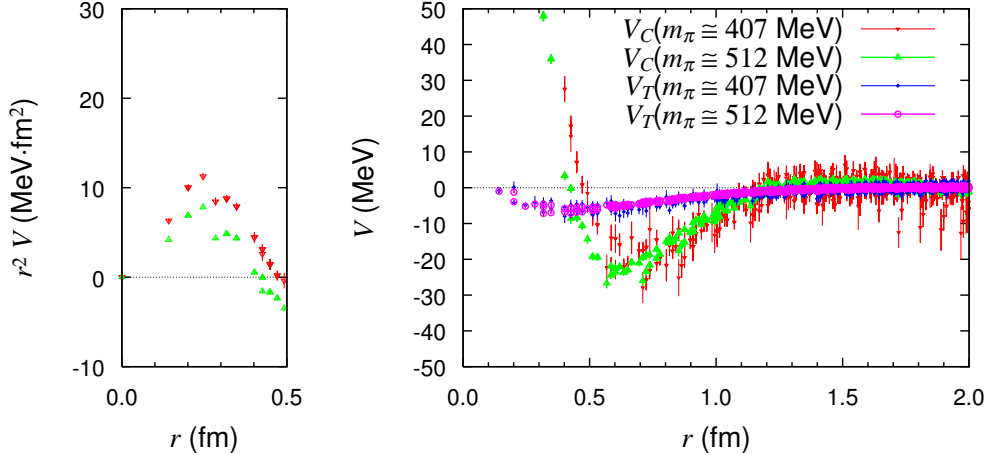}%
 \caption{
  The central and the tensor potentials in $^3S_1-^3D_1$ channel in quenched QCD
  at  $m_\pi\simeq 407$ MeV (red and blue) and 512 MeV (green and magenta).
 }
 \label{BGL.quenchL32_VCT_3E1}
\end{figure}
\begin{figure}[h]
 \centering \leavevmode
  \includegraphics[width=.90\textwidth]
  {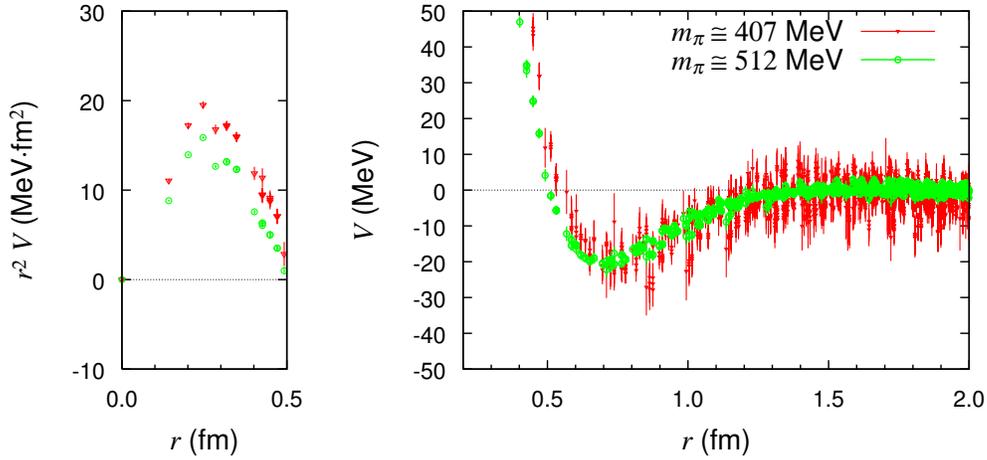}%
 \caption{
 The central potential in $^1S_0$ channel in quenched QCD  at $m_\pi\simeq 407$ MeV (red) and 512 MeV (green).
 }
 \label{BGL.quenchL32_VCeff_1S0}
\end{figure}

\subsection{Quenched QCD}

Figures~\ref{BGL.quenchL32_VCT_3E1} and~\ref{BGL.quenchL32_VCeff_1S0} 
show the $\Lambda N$ potentials 
with $(m_\pi,m_K) \approx (512,606)$ and $(407,565)$ MeV,
in quenched QCD at the time-slice $t-t_0=7$.
Fig.~\ref{BGL.quenchL32_VCT_3E1} shows the $V_C(J=1)$ and $V_T$, 
while the Fig.~\ref{BGL.quenchL32_VCeff_1S0} shows the $V_C(J=0)$. 
We find that %
the qualitative behaviors of the $\Lambda N$
potential in quenched QCD are more or less similar to those in full QCD
in both  $J=1$ and $0$ channels:
Namely, the attractive pocket of the central potential 
moves to longer distance region as the quark mass decreases, and
the quark mass dependence of the
tensor potential seems to be small.

\subsection{Scattering lengths}

\begin{figure}[h]
 \centering \leavevmode
  \includegraphics[width=.56\textwidth]
  {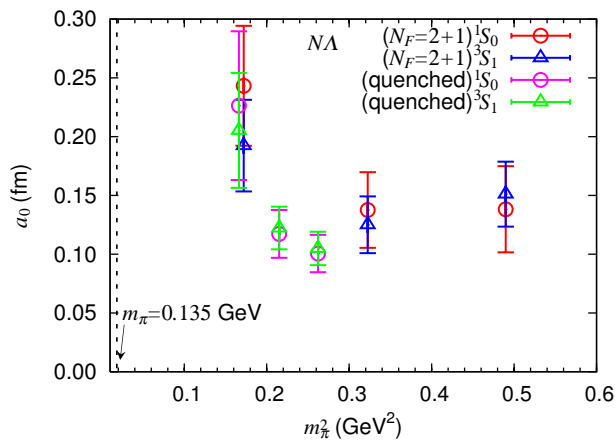}%
 \caption{
 Scattering lengths of $\Lambda N$ interaction as a function of $m_\pi^2$.}
 \label{sctlngNL_vs_pimasssq_marked}
\end{figure}

Figure~\ref{sctlngNL_vs_pimasssq_marked} shows 
the scattering lengths as a function of $m_\pi^2$, 
which are calculated through the L\"{u}scher's
formula \cite{Luscher,CPPACS.PRD71_94504_2005}
\begin{equation}
 \begin{array}{l}
  k \cot \delta_0(k)
   = {2\over \sqrt{\pi}L}Z_{00}(1;(kL/(2\pi))^2)
   = 1/a_0 + O(k^2), \\
  \mbox{with} \quad 
   Z_{00}(s;q^2)={1\over \sqrt{4\pi}}\sum_{\vec{n}\in \bfi{Z}^3}
   (n^2-q^2)^{-s} \quad ({\rm Re}\ s> 3/2), 
 \end{array}
\end{equation}
where $Z_{00}(1;q^2)$ is obtained by the analytic continuation in $s$. 
The 
energy, $E={k^2\over 2\mu}$, on the finite lattice volume is 
determined by fitting the asymptotic region of the BS wave function 
in terms of the Green's function
\begin{equation}
 G(\vec{r},k^2)={1\over L^3}\sum_{\vec{p}\in\Gamma}{1\over p^2-k^2}
  {\rm e}^{i\vec{p}\cdot\vec{r}},
  \qquad
  \Gamma = \left\{
	    \vec{p}, \vec{p} = \vec{n}{2\pi\over L},
	    \vec{n} \in {\bfi{Z}}^3
	   \right\},
\end{equation}
which is the solution to the Helmholtz equation 
$(\Delta+k^2)G(\vec{r},k^2)=-\delta_L(\vec{r})$ with $\delta_L(\vec{r})$
being the periodic delta function\cite{Luscher,CPPACS.PRD71_94504_2005}.

As is seen in the Fig.~\ref{sctlngNL_vs_pimasssq_marked}, 
the scattering lengths are almost constant for larger $u,d$ 
quark mass corresponding to $m_\pi \gsim 560$ MeV. 
On the other hand, for lighter $u,d$ quark mass region that
$400$ MeV $\lsim m_\pi \lsim$ 500 MeV, the present 
result seems to show that the scattering lengths increase 
as the $u,d$ quark mass decreases. 
The present values are still much smaller than 
the empirical scattering lengths of $\Lambda N$,
 $a_0 \sim 1.5 - 2.5$ fm,
 estimated  from the measurement of the 
 $\Lambda N$ total cross section and 
the theoretical studies of $\Lambda$-hypernuclei. 
The scattering lengths calculated  in quenched QCD are
qualitatively similar  to those in  $2+1$ flavor QCD. 
As is discussed in the former subsection, 
we will need larger spatial volume in full QCD to 
extract the scattering lengths reliably at $m_\pi \lsim 300$ MeV.

\section{Summary}

We have calculated the central and tensor parts of $\Lambda N$ potentials 
as well as the scattering lengths in lattice QCD at several values of 
the $u,d$ quark mass corresponding to
$m_\pi \simeq 300 - 700$ MeV. 
The central potentials indicate that the interaction range 
becomes larger 
while the depth of the attractive well hardly changes 
as the $u,d$ quark mass decreases. 
On the other hand the tensor force has relatively a weak quark mass dependence. 
The present result of the scattering lengths %
shows that 
the $\Lambda N$ interaction is attractive and becomes stronger 
as the $u,d$ quark mass decreases. 
The calculation in $2+1$ full QCD
for the $\Lambda N$ system 
with larger volume and smaller lattice spacing 
at physical quark mass 
is highly 
desirable for definite conclusions on the 
nature of the $\Lambda N$ interaction.

\acknowledgments

{%
The authors would 
like to thank  PACS-CS Collaboration for 
allowing us to access their full QCD gauge configurations, 
and 
Dr.~T.~Izubuchi 
for providing a sample FFT code.
}
The full QCD calculations have been done 
by using PACS-CS computer 
under the ``Interdisciplinary Computational Science Program'' 
of Center for Computational Science, University of Tsukuba 
(No 09a-11).
The quenched QCD calculations have been done 
by using Blue Gene/L computer 
under the ``Large scale simulation program''
at KEK (No. 09-23).
H.N. would also like to thank Dr.~K.~Itahashi 
and 
Advanced Meson Science Laboratory of RIKEN Nishina Center 
for providing a special computer resource. 
H.~N. is supported by the Global COE Program for 
Young Researchers at Tohoku University (No. 22210005). 
This research was partly supported 
by the MEXT Grant-in-Aid,  Scientific Research on Priority Areas
(No. 20028013) and Scientific Research on Innovative Areas 
(Nos. 21105515, 20105003).

\end{document}